\documentclass[pra,onecolumn,showpacs]{revtex4}
\usepackage{amssymb}
\usepackage{amsmath}
\usepackage{graphicx}

\begin{document}

\title{Interactions of solitons with positive and negative masses: Shuttle
motion and co-acceleration}
\author{Hidetsugu Sakaguchi$^{1}$ and Boris A. Malomed$^{2}$ }
\address{$^{1}$Department of Applied Science for Electronics and Materials,
Interdisciplinary Graduate School of Engineering Sciences, Kyushu
University, Kasuga, Fukuoka 816-8580, Japan \\
$^{2}$Department of
Physical Electronics, School of Electrical Engineering, Faculty of
Engineering, and Center for Light-Matter Interaction, Tel Aviv
University, Tel Aviv 69978, Israel}

\begin{abstract}
We consider a possibility to realize self-accelerating motion of interacting
states with effective positive and negative masses in the form of pairs of
solitons in two-component BEC loaded in an optical-lattice (OL) potential. A
crucial role is played by the fact that gap solitons may feature a negative
dynamical mass, keeping their mobility in the OL. First, the respective
system of coupled Gross-Pitaevskii equations (GPE) is reduced to a system of
equations for envelopes of the lattice wave functions. Two generic dynamical
regimes are revealed by simulations of the reduced system, \textit{viz}%
.,shuttle oscillations of pairs of solitons with positive and negative
masses, and splitting of the pair. The co-accelerating motion of the
interacting solitons, which keeps constant separation between them, occurs
at the boundary between the shuttle motion and splitting. The position of
the co-acceleration regime in the system's parameter space can be adjusted
with the help of an additional gravity potential, which induces its own
acceleration, that may offset the relative acceleration of the two solitons,
while gravity masses of both solitons remain positive. The numerical
findings are accurately reproduced by a variational approximation.
Collisions between shuttling or co-accelerating soliton pairs do not alter
the character of the dynamical regime. Finally, regimes of the shuttle
motion, co-acceleration, and splitting are corroborated by simulations of
the original GPE system, with the explicitly present OL potential.
\end{abstract}

\maketitle

\section{Introduction}

Search for robust self-accelerating pulses in various physical settings has
drawn much interest, starting from the discovery of Airy-wave modes in
quantum mechanics \cite{Berry}. Experimentally, this propagation mode was
demonstrated in quantum matter represented by electron beams (under
conditions which make interactions between electrons negligible) \cite%
{electron}. Using the similarity of the linear Schr\"{o}dinger equation for
the wave function of quantum particles to the paraxial wave-propagation
equation in classical-field systems, the realization of Airy waves was
elaborated in optics \cite{siv}, plasmonics \cite{plasmonics}, gas discharge
\cite{gas-discharge}, acoustics \cite{acc}, and hydrodynamics \cite{water}.
Further, the commonly known similarity of the Schr\"{o}dinger equation to
the Gross-Pitaevskii equation (GPE) for the mean-field wave function of
atomic Bose-Einstein condensates (BECs) makes it possible to predict
Airy-wave modes in atomic BEC as well \cite{kli}.

Full Airy waves carry an infinite norm (alias diverging integral power, in
terms of optics), therefore truncated waves with a finite norm were used in
the theory and experiments \cite{siv,pan}, although the truncation leads to
gradual destruction of the self-accelerating wave pattern. The medium's
nonlinearity may also be detrimental to the evolution of the Airy waves,
which are introduced as eigenmodes of the linear propagation \cite{pan}-\cite%
{Thawatchai}.

For these reasons, a relevant objective is to design physical models that
would allow self-accelerated propagation of well-localized modes with a
finite norm, which would be maintained by the nonlinearity, rather than
being damaged by it. Actually, this objective implies looking for models
that should support stable self-acceleration of quasi-soliton states. In
particular, this possibility was recently predicted for one- and
two-dimensional hybrid (matter-wave -- microwave) solitons produced by the
interplay of a two-component BEC\ and a resonant electromagnetic field which
couples the components \cite{Qin}. Another approach relies on the well-known
idea that a pair of objects with positive and negative masses may develop
constant self-acceleration under the action of interaction forces \cite%
{neg-mass}. While real bodies with a negative mass do not exist,
quasi-particles and wave pulses may acquire an effective negative mass in
various settings. In this direction, as essential result was the prediction
\cite{Peschel} and experimental realization \cite{Peschel-exper} of bound
pulses in nonlinear photonic crystals with opposite signs of the dispersion
(effective mass) of their two components. Theoretically, a similar result
was predicted for a pair of correlated quantum particles coupled by
long-range interaction, which perform hopping in a Bose-Hubbard lattice, as
one of the particles may also acquire an effective negative mass in the
lattice \cite{Longhi}.

The objective of the present work is to explore a possibility of forming
bound states of solitons with opposite signs of the effective masses, which
implies that they should also have opposite signs of the self-interaction
coefficients (otherwise, bright solitons cannot exist in both components;
for this reason, only one component was a soliton in the above-mentioned
photonic setting \cite{Peschel}, while the other one was treated as a
Thomas-Fermi mode). This situation is possible in a two-component atomic BEC
loaded in an optical-lattice (OL) potential, which may induce the effective
mass of either sign (positive for regular solitons, and negative for gap
solitons in a finite bandgap \cite{KonSal}-\cite{Yulik2}), while the sign of
the self-interaction in any component may be switched by means of the
Feshbach resonance \cite{FR}. It is relevant to mention that the dynamics of
a pair of matter-wave solitons with effective masses of opposite signs,
loaded in a harmonic-oscillator trapping potential, was studied in recent
work \cite{Yulik2}. As a result, the soliton with the positive mass remains
trapped, while its counterpart with the negative mass can escape, as the
potential is effectively expulsive for it \cite{we}.

The model is introduced in Section II, first in the form of the nonlinearly
coupled GPEs with spatially periodic potentials representing the OL \cite%
{gapsol}. Then, we apply the approximation of slowly-varying envelope
amplitudes to derive a free-space GPE system with opposite effective masses
and opposite signs of the self-interaction. In Section III, simulations of
the latter system demonstrate that it gives rise to two generic dynamical
regimes: spontaneous shuttle oscillations of pairs of interacting solitons,
with the separation between them also oscillating (so that the solitons
periodically pass through each other), and splitting of the pair. The
co-acceleration of the positive- and negative-mass solitons, which keep a
constant distance between themselves, takes place at the boundary between
the two regimes (a similar observation suggesting the non-generic character
of the co-accelerating motion of the pair of interacting pulses, one of
which was not a soliton, was reported in Ref. \cite{Peschel}; however, the
generic regime of the shuttle motion was not reported in that work). Note
that the shuttle regime also implies that the two soliton stay paired and
spontaneously develop common acceleration, but with a periodically reversing
sign. Also in Section III, we develop a variational approximation (VA),
which accurately predicts the shuttle, co-acceleration, and splitting
regimes. Further, in the same section we consider the system which
additionally includes a gravity potential (it is important to note that,
while the effective dynamical mass of one soliton is negative, its gravity
mass remains normal positive). Using the fact that the gravity also imparts
acceleration to the solitons, we demonstrate, both numerically and by means
of the VA, that the gravity-induced acceleration can offset the splitting
force, and thus adjust the location of the co-acceleration regime in the
system's parameter space. In addition, we report results of simulation of
collisions between soliton pairs, in both the shuttle and co-acceleration
regimes, the result being that the collisions may change the separation
between the paired solitons, but not the character of the dynamical regime.
Finally, in Section IV we return to the underlying system of GPEs which
explicitly includes the OL potential, and demonstrate, by means of
systematic simulations, that the same regimes, \textit{viz}., the shuttle
motion, co-acceleration, and splitting, are produced by that system,
including its extended version with the gravity potential. The paper is
concluded by Section V.

\section{Models: the optical lattice and slowly varying envelopes}

We start with the system of scaled GPEs for a binary BEC, with equal atomic
masses of its two components, $\phi $ and $\psi $, loaded in the OL
potential, whose period is scaled to be $1$, with amplitudes $-U_{1,2}$ \cite%
{gapsol}:
\begin{eqnarray}
i\frac{\partial \phi }{\partial t} &=&-\frac{1}{2}\frac{\partial ^{2}\phi }{%
\partial x^{2}}-\left[ g_{1}|\phi |^{2}+\gamma |\psi |^{2}+U_{1}\cos \left(
2\pi x\right) \right] \phi ,  \notag \\
i\frac{\partial \psi }{\partial t} &=&-\frac{1}{2}\frac{\partial ^{2}\psi }{%
\partial x^{2}}-\left[ \gamma |\phi |^{2}-g_{2}|\psi |^{2}+U_{2}\cos \left(
2\pi x\right) \right] \psi .  \label{cos}
\end{eqnarray}%
Here, $g_{1}>0$ and $-g_{2}<0$ are coefficients of the self-interaction of
the components, implying that, as said above, their signs are made opposite
by means of the Feshbach resonance applied to one of the components, and $%
\gamma >0$ is the coefficient of the cross-attraction. It is well known that
the GPEs, based on the mean-field approximation, provide a very accurate
model of the atomic BEC. The only exception occurs in the case of a binary
atomic BEC, when the self-repulsion in both components almost exactly
cancels with attraction between them, making the beyond-mean-field terms,
generated by quantum fluctuations, important corrections to the GPE system
\cite{Petrov}. This is definitely not the case in the present setting.

To focus on the case of opposite signs of the effective mass for solitons in
the interacting components, $\phi $ and $\psi $, we consider the case when
quasi-wavenumbers of wave functions $\phi $ and $\psi $ are set to be close,
respectively, to the center and edge of the first OL's Brillouin zone, in
terms of Eq. (\ref{cos}). Near the center, which corresponds to the zero
quasi-wavenumber, the effective mass, calculated by means of the known
methods \cite{Pu,we,gapsol}, is
\begin{equation}
M_{1}=\frac{2\pi ^{3}+U_{1}^{2}+\pi ^{2}\sqrt{4\pi ^{4}+2U_{1}^{2}}}{10\pi
^{4}+U_{1}^{2}-3\pi ^{2}\sqrt{4\pi ^{4}+2U_{1}^{2}}},  \label{meff1}
\end{equation}%
and the wave function itself is approximated as
\begin{gather}
\phi (x)=\Phi (x)\frac{1+2a\cos \left( 2\pi x\right) }{\sqrt{1+2a^{2}}},
\label{phi} \\
a\equiv \sqrt{\left( \frac{\pi ^{2}}{U_{1}}\right) ^{2}+\frac{1}{2}}-\frac{%
\pi ^{2}}{U_{1}},  \label{a}
\end{gather}%
where $\Phi (x)$ is the slowly varying envelope amplitude. Near the edge of
the Brillouin zone, which corresponds to quasi-wavenumber $\pi $, the
effective mass is
\begin{equation}
-M_{2}=\frac{U_{2}}{U_{2}-2\pi ^{2}}  \label{meff2}
\end{equation}%
(it is defined with sign minus, to focus below on the relevant case of the
negative mass, $M_{2}>0$), with the respective wave function
\begin{equation}
\psi (x)=\sqrt{2}\Psi (x)\cos \left( \pi x\right)   \label{psi}
\end{equation}%
and slowly varying envelope amplitude $\Psi (x)$. The slow variation implies
that solitons represented by $\Phi $ and $\Psi $ may be relevant solutions
if their width $l$ is much larger than periods of spatial oscillations of
the carrier wave functions (\ref{phi}) and (\ref{psi}), i.e.,
\begin{equation}
l\gg 1.  \label{>>}
\end{equation}

The substitution of expressions (\ref{phi}) and (\ref{psi}) into original
equations (\ref{cos}) leads, by means of the procedure of averaging with
respect to rapid oscillations of the carrier wave functions \cite{we,gapsol}%
, to equations governing the slow evolution of the envelope amplitudes,
which do not include an external potential:
\begin{eqnarray}
i\frac{\partial \Phi }{\partial t} &=&-\frac{1}{2M_{1}}\frac{\partial
^{2}\Phi }{\partial x^{2}}-\left( G_{1}|\Phi |^{2}+\Gamma |\Psi |^{2}\right)
\Phi ,  \label{PHI} \\
i\frac{\partial \Psi }{\partial t} &=&\frac{1}{2M_{2}}\frac{\partial
^{2}\Psi }{\partial x^{2}}-\left( \Gamma |\Phi |^{2}-G_{2}|\Psi |^{2}\right)
\Psi ,  \label{PSI}
\end{eqnarray}%
with effective nonlinearity coefficients,
\begin{gather}
G_{1}=g_{1}\frac{1+12a^{2}+6a^{4}}{(1+2a^{2})^{2}},G_{2}=\frac{3}{2}g_{2},
\notag \\
\Gamma =\frac{1+2a^{2}+2a}{1+2a^{2}}.  \label{geff}
\end{gather}%
Numerical results are reported below both for the reduced system of Eqs. (%
\ref{PHI}) and (\ref{PSI}) (in Section III), and for the underlying one,
based on Eq. (\ref{cos}) (in Section IV).

Sign minus is eliminated in front of the second derivative in Eq. (\ref{PSI}%
) according to the definition of the respective effective mass in Eq. (\ref%
{meff2}). Accordingly, it is obvious that Eqs. (\ref{PHI}) and (\ref{PSI})
may indeed feature opposite signs of the effective masses, if $M_{1}$ and $%
M_{2}$ are both positive (or both negative), and opposite signs of the
effective coefficients of the self-interaction in the two components, if $%
G_{1}$ and $G_{2}$ are both positive (or both negative) too. These sign
combinations open the way to the creation of pairs of bright solitons with
opposite signs of their dynamical masses, which is the objective outlined in
the introduction. We also fix $\Gamma >0$, although the sign of this
coefficient can be reversed by a combination of the complex conjugation and
swap $\Phi \rightleftarrows \Psi $. Note that Eqs. (\ref{PHI}) and (\ref{PSI}%
) keep the Galilean invariance, in spite of opposite signs of the mass
parameters in them, therefore it is easy to find soliton complexes moving
with an arbitrary velocity, as shown below.

\section{Dynamics of paired envelope solitons: numerical and analytical
results}

\subsection{Exact solutions for soliton complexes}

In the basic case of $M_{1,2}>0$ and $G_{1,2}>0$, Eqs. (\ref{PHI}) and (\ref%
{PSI}) generate, in addition to obvious single-component solitons, a family
of exact steady-state soliton complexes with free parameter $\kappa $ (the
inverse width) and an arbitrary velocity, $v$:
\begin{gather}
\Phi =\frac{A\exp \left[ iM_{1}vx-\frac{i}{2}\left( \frac{\kappa ^{2}}{M_{1}}%
+M_{1}v^{2}\right) t\right] }{\cosh \left( \kappa \left( x-vt\right) \right)
},  \label{solPhi} \\
\Psi =\frac{B\exp \left[ -iM_{2}vx+\frac{i}{2}\left( \frac{\kappa ^{2}}{M_{2}%
}+M_{2}v^{2}\right) t\right] }{\cosh \left( \kappa \left( x-vt\right)
\right) },  \label{solPsi} \\
A^{2}=\frac{\kappa ^{2}}{M_{1}M_{2}}\frac{M_{2}G_{2}-M_{1}\Gamma }{%
G_{1}G_{2}+\Gamma ^{2}},  \label{A} \\
B^{2}=\frac{\kappa ^{2}}{M_{1}M_{2}}\frac{M_{1}G_{1}+M_{2}\Gamma }{%
G_{1}G_{2}+\Gamma ^{2}},  \label{B}
\end{gather}%
provided that expressions (\ref{A}) and (\ref{B}) take positive values. It
is relevant to mention that this solution represents only a particular case
of a more general family of stationary two-component solitons, as, in the
case of $v=0$, the generic soliton solution must feature two independent
parameters, which may be defined as norms of the two components,%
\begin{equation}
N_{1,2}=\int_{-\infty }^{+\infty }\left\vert \Phi (x),\Psi (x)\right\vert
^{2}dx,  \label{N}
\end{equation}%
while the exact solution (\ref{solPhi}), (\ref{solPsi}) contains only one
free parameter, $\kappa $, at $v=0$. If Eqs. (\ref{PHI}) and (\ref{PSI}) are
derived from the underlying GPEs by means of the above-mentioned averaging
procedure, the corresponding condition (\ref{>>}) amounts to restriction $%
\kappa \ll 1$. However, the scaling invariance of Eqs. (\ref{PHI}) and (\ref%
{PSI}) implies that simulations of the equations may be actually performed
for $\kappa =1$ (as it is done below), and the results can be then rescaled
for any other value of $\kappa $.

In accordance with what is reported below for soliton pairs with a finite
separation between their constituents, the steady-state complexes are stable
under condition $\beta >0$ imposed on the two components, see Eq. (\ref{beta}%
) below. In the opposite case of $\beta <0$, the complexes are unstable
against splitting into separating components.

\subsection{Initial numerical results: shuttle and self-accelerating motion
of paired solitons}

The soliton complexes given by Eqs. (\ref{solPhi})-(\ref{B}) do not feature
self-acceleration, being built of two components which are located at the
bottom of the potential of their mutual attraction, hence no interaction
forces act on them. As said above, our main objective is to look for
self-accelerating soliton pairs. This may be possible if the constituents
are separated by some distance, which gives rise to opposite interaction
forces applied to them. Acting on the solitons with opposite signs of the
dynamical mass, these forces should produce accelerations with identical
signs.

\begin{figure}[h]
\begin{center}
\includegraphics[height=4.cm]{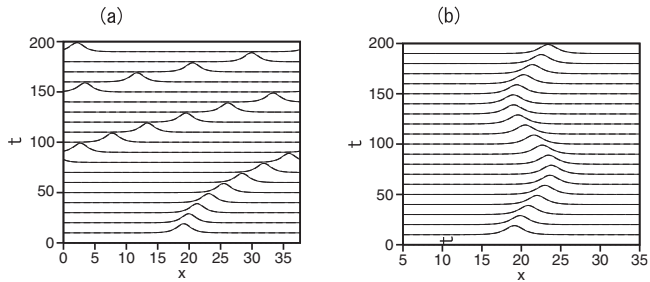}
\end{center}
\caption{The evolution of $\left\vert \Phi \left( x,t\right) \right\vert $
and $\left\vert \Psi \left( x,t\right) \right\vert $ (solid and dashed
lines, respectively, which nearly overlap) at $M_{2}^{-1}=1$ (a) and $%
M_{2}^{-1}=0.96$ (b), with other parameters and the input taken as per Eqs. (%
\protect\ref{params}) (\protect\ref{input}), respectively.}
\label{fig1}
\end{figure}

To realize this possibility, we started simulations of Eqs. (\ref{PHI}) and (%
\ref{PSI}) for the soliton complexes given by Eqs. (\ref{solPhi}) and (\ref%
{solPsi}) with $\kappa =1$ and $v=0$, choosing other parameters as%
\begin{equation}
M_{1}=1,G_{1}=0.9,\Gamma =0.1,G_{2}=0.1+M_{2}^{-1},  \label{params}
\end{equation}%
while $M_{2}^{-1}$ will be varied as a control parameter. In this case, Eqs.
(\ref{A}) and (\ref{B}) yield $A=B=1$. The separation between the
constituents, $x_{0}$, which is necessary to introduce the interaction
forces, was introduced by taking the initial conditions as
\begin{equation}
\Phi _{0}=\mathrm{sech~}x,\Psi _{0}=\mathrm{sech}\left( x-x_{0}\right)
,~x_{0}=0.1,  \label{input}
\end{equation}%
whose norms (\ref{N}) are $N_{1}=N_{2}=2$. The simulations were performed
with periodic boundary conditions, taking the period which is much larger
than widths of the produced solitons, as seen in Fig. \ref{fig1}(a).

\begin{figure}[h]
\begin{center}
\includegraphics[height=4.cm]{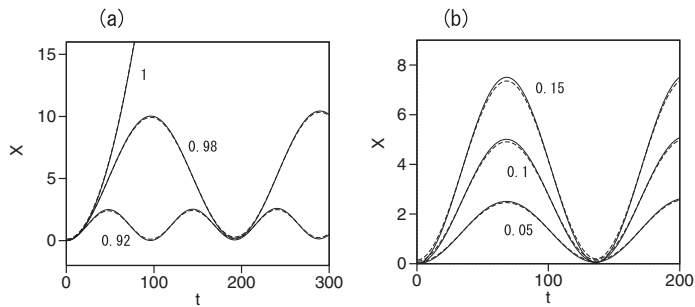}
\end{center}
\caption{Trajectories of centers of the $\Phi $ and $\Psi $ components
(solid and dashed lines, respectively): (a) at $M_{2}^{-1}=0.92,0.98,$ and $1
$, for $x_{0}=0.1$ in Eq. (\protect\ref{input}); and (b) at $x_{0}=0.05$, $%
0.10$, and $0.15$, for fixed $M_{2}^{-1}=0.96$.}
\label{fig2}
\end{figure}

Figure \ref{fig1} shows the evolution of the wave functions, in terms of $%
|\Phi \left( x,t\right) |$ and $|\Psi \left( x,t\right) |$, at (a) $%
M_{2}^{-1}=1$ and (b) $M_{2}^{-1}=0.96$. Permanent self-acceleration of the
bound soliton pair is observed in Fig. \ref{fig1}(a). However, this regime
of motion is not a generic one, in terms of varying control parameter $%
M_{2}^{-1}$ (a conclusion that such a regime is not generic was also made in
Ref. \cite{Peschel}): as seen in Fig. \ref{fig2}(b), the interacting
solitons exhibit shuttle motion, with periodically sign-changing
co-acceleration, at $M_{2}^{-1}<1$. The shuttle period diverges at $%
M_{2}^{-1}\rightarrow 1$, and the solitons separate at $M_{2}^{-1}>1$. These
conclusions are confirmed by Fig. \ref{fig2}(a), which displays trajectories
of the motion of centers of both constituent solitons at $M_{2}^{-1}\leq 1$.
Collecting results of simulations carried out at other values of the
parameters suggests that, in the general case, the permanent co-acceleration
occurs under a balance condition,%
\begin{equation}
\beta \equiv \frac{N_{2}}{M_{1}}-\frac{N_{1}}{M_{2}}=0,  \label{beta}
\end{equation}%
which is derived below analytically by means of the VA.

Fixing the parameters as per Eq. (\ref{params}) and $M_{2}^{-1}=0.96$, the
initial separation $x_{0}$ between the constituent solitons in Eq. (\ref%
{input}) is varied in Fig. \ref{fig2}(b). It is observed that the shuttle
motion persists in this case, with the amplitude growing proportionally to $%
x_{0}$ (the same result is derived below by means of the VA). On the other
hand, the two solitons separate in the case of $M_{2}^{-1}>1$.

Thus, the co-acceleration regime plays the role of a separatrix between two
generic regimes of motion, \textit{viz}., the shuttle oscillations and
splitting of the soliton pair. These conclusions, suggested by the
systematic simulations, are explained by means of the VA developed below.

\subsection{The variational approximation (VA)\ and comparison with
numerical results}

The system of Eqs. (\ref{PHI}) and (\ref{PSI})\ for envelope wave functions
can be derived from the Lagrangian,%
\begin{gather}
L=\int_{-\infty }^{+\infty }dx\left\{ \frac{i}{2}\left( \frac{\partial \Phi
}{\partial t}\Phi ^{\ast }-\frac{\partial \Phi ^{\ast }}{\partial t}\Phi +%
\frac{\partial \Psi }{\partial t}\Psi ^{\ast }-\frac{\partial \Psi ^{\ast }}{%
\partial t}\Psi \right) \right.   \notag \\
\left. -\frac{1}{2M_{1}}\left\vert \frac{\partial \Phi }{\partial x}%
\right\vert ^{2}+\frac{1}{2M_{2}}\left\vert \frac{\partial \Psi }{\partial x}%
\right\vert ^{2}+\left( \frac{G_{1}}{2}|\Phi |^{4}-\frac{G_{2}}{2}|\Psi
|^{4}+\Gamma |\Phi |^{2}|\Psi |^{2}\right) \right\} .  \label{L}
\end{gather}%
The solitons with amplitudes $A_{1,2}$, coordinates $\xi _{1,2}$, momenta $%
k_{1,2}$, and overall phases $\varphi _{1,2}$ may be approximated by the
usual Gaussian ansatz \cite{Anderson,Progress}:
\begin{equation}
\left\{ \Phi ,\Psi \right\} =A_{1,2}\exp \left[ i\varphi _{1,2}(t)-\alpha
_{1,2}\left( x-\xi _{1,2}(t)\right) ^{2}+ik_{1,2}(t)\left( x-\xi
_{1,2}\right) \right] .  \label{ansatz}
\end{equation}%
The substitution of the ansatz in Eq. (\ref{L}) leads to the effective
Lagrangian,
\begin{gather}
L_{\mathrm{eff}}=-N_{1}\dot{\varphi}_{1}-N_{2}\dot{\varphi}_{2}+\frac{N_{1}}{%
2M_{1}}\alpha _{1}+\frac{N_{2}}{2M_{2}}\alpha _{2}  \notag \\
+\frac{G_{1}}{2}\sqrt{\frac{\alpha _{1}}{\pi }}N_{1}^{2}-\frac{G_{2}}{2}%
\sqrt{\frac{\alpha _{2}}{\pi }}N_{2}^{2}  \notag \\
+\Gamma \sqrt{\frac{2\alpha _{1}\alpha _{2}}{\pi (\alpha _{1}+\alpha _{2})}}%
N_{1}N_{2}\exp \left[ -\frac{2\alpha _{1}\alpha _{2}}{\alpha _{1}+\alpha _{2}%
}(\xi _{1}-\xi _{2})^{2}\right]   \notag \\
-\frac{N_{1}}{2M_{1}}k_{1}^{2}+\frac{N_{2}}{2M_{2}}k_{2}^{2}+N_{1}k_{1}\dot{%
\xi _{1}}+N_{2}k_{2}\dot{\xi _{2}},  \label{Leff}
\end{gather}%
where the overdot stands for $d/dt$, and amplitudes $A_{1,2}$ are expressed
in terms of the respective norms, $N_{1}=\sqrt{\pi /(2\alpha _{1})}A_{1}^{2}$
and $N_{2}=\sqrt{\pi /(2\alpha _{2})}A_{2}^{2}$. Being dynamical invariants
of the system, the norms are treated as constants. The Lagrangian gives rise
to the system of the Euler-Lagrange equations, which, upon the elimination
of $\dot{k}_{1,2}$, can be cast in the form of two coupled second-order
equations of motion for coordinates $\xi _{1,2}$ (unessential equations for $%
\dot{\varphi}_{1,2}$ are not written here):
\begin{equation}
\frac{d^{2}\xi _{1,2}}{dt^{2}}=\frac{N_{2,1}}{M_{1,2}}\alpha \exp \left[ -%
\frac{2\alpha _{1}\alpha _{2}}{\alpha _{1}+\alpha _{2}}(\xi _{1}-\xi
_{2})^{2}\right] (\xi _{2}-\xi _{1}),  \label{xi1}
\end{equation}%
with
\begin{equation}
\alpha \equiv \frac{2\Gamma }{\sqrt{\pi }}\left( \frac{2\alpha _{1}\alpha
_{2}}{\alpha _{1}+\alpha _{2}}\right) ^{3/2}.  \label{alpha}
\end{equation}%
It is seen that the right-hand sides of Eq. (\ref{xi1}) for $\xi _{1}$ and $%
\xi _{2}$ have identical signs in the case of $M_{1}M_{2}>0$ (recall we are
dealing with the case of $M_{1,2}>0$), which indeed implies the
co-acceleration of the solitons. That is, the first soliton, denoted by $\xi
_{1}$, is attracted to the second soliton denoted by $\xi _{2}$, while the
latter one is repelled from the first soliton. On the other hand, the total
momentum of the soliton pair is, as follows from Lagrangian (\ref{Leff}),%
\begin{equation}
P=N_{1}M_{1}\frac{d\xi _{1}}{dt}-N_{2}M_{2}\frac{d\xi _{2}}{dt},
\end{equation}%
and it immediately follows from Eq. (\ref{xi1}) that $P$ remains a dynamical
invariant, even if the pair as a whole is moving with acceleration.

It is straightforward to combine equations (\ref{xi1}) for $\xi _{1}$ and $%
\xi _{2}$, deriving an equation for separation $\Delta \xi =\xi _{2}-\xi _{1}
$ between the solitons:
\begin{equation}
\frac{d^{2}\Delta \xi }{dt^{2}}=-\alpha \beta \exp \left[ -\frac{2\alpha
_{1}\alpha _{2}}{\alpha _{1}+\alpha _{2}}(\Delta \xi )^{2}\right] \Delta \xi
,  \label{Delta_xi}
\end{equation}%
where $\beta $ is defined as per Eq. (\ref{beta}). In particular, for small $%
\left\vert \Delta \xi \right\vert $ the linearization of Eq. (\ref{Delta_xi}%
) yields%
\begin{equation}
\frac{d^{2}\Delta \xi }{dt^{2}}=-\alpha \beta \Delta \xi ~,  \label{linear}
\end{equation}%
It follows from Eq. (\ref{linear}) that, in the case of $\alpha \beta >0$
[i.e., $\Gamma \beta >0$, as it follows from Eq. (\ref{alpha})], the
separation between the interacting solitons preforms periodic oscillations
with arbitrary amplitude $x_{0}$,
\begin{equation}
\Delta \xi =x_{0}\cos \left( \sqrt{\alpha \beta }t\right) ,  \label{osc}
\end{equation}%
while in the opposite case, $\alpha \beta >0$, the separation monotonously
grows in time, i.e., the interacting solitons separate. The latter
analytical result provides a direct explanation to the separation regime
revealed above by the numerical simulations of Eqs. (\ref{PHI}) and (\ref%
{PSI}) at $\beta <0$.

Exactly at $\beta =0$, separation $\Delta \xi $ keeps the initial value, $%
x_{0}$, hence Eq. (\ref{xi1}) predicts permanent co-acceleration of the
paired solitons, with the acceleration itself proportional to the initial
separation, $x_{0}$. This analytical result explains the most essential
numerical finding reported above: the co-accelerating motion of the
internally stationary soliton pair at $\beta =0$.

To address the shuttle motion revealed by the simulations in Figs. \ref{fig1}
and \ref{fig2} at $\beta >0$, we note that the linearized version of Eq. (%
\ref{xi1}) gives rise to the following equation of motion for the mean
position of the pair, $\ \Xi \equiv (\xi _{1}+\xi _{2})/2$:
\begin{equation}
\frac{d^{2}\Xi }{dt^{2}}=\frac{\alpha }{2}\left( \frac{N_{2}}{M_{1}}+\frac{%
N_{1}}{M_{2}}\right) \Delta \xi =\frac{\alpha }{2}\left( \frac{N_{2}}{M_{1}}+%
\frac{N_{1}}{M_{2}}\right) x_{0}\cos \left( \sqrt{\beta \alpha }t\right) ,
\label{Xi}
\end{equation}%
where solutions (\ref{osc}) for $\Delta \xi $ is substituted. Then, the
solution to Eq.~(\ref{Xi}) is%
\begin{eqnarray}
\Xi &=&Rx_{0}\left[ 1-\cos \left( \sqrt{\beta \alpha }t\right) \right] ,
\label{X(t)} \\
R &\equiv &\left( 2\beta \right) ^{-1}\left( \frac{N_{2}}{M_{1}}+\frac{N_{1}%
}{M_{2}}\right) ,  \label{R}
\end{eqnarray}%
if the initial value of $\Xi $ and overall velocity are zero. This result
explains the shuttle motion of the soliton pair observed in Figs. \ref{fig1}
and \ref{fig2}, as well as the above-mentioned fact, also revealed by the
direct simulations, that the amplitude of the shuttle oscillations grows
proportionally to $x_{0}$. Further, in the limit of $\beta \rightarrow 0$,
Eq. (\ref{Xi}) precisely reproduces the permanent co-acceleration of the
pair, which was revealed by the direct simulations close to $\beta =0$:%
\begin{equation}
\Xi \left( \beta =0\right) =\frac{1}{2}\mathrm{a}t^{2},~\mathrm{a}\equiv
\frac{1}{2}\left( \frac{N_{2}}{M_{1}}+\frac{N_{1}}{M_{2}}\right) \alpha
x_{0}.  \label{co-accel}
\end{equation}%
If full equation (\ref{xi1}) is used, without the linearization, the
acceleration is%
\begin{equation}
\mathrm{a}=\frac{1}{2}\left( \frac{N_{2}}{M_{1}}+\frac{N_{1}}{M_{2}}\right)
\alpha \exp \left( -\frac{2\alpha _{1}\alpha _{2}}{\alpha _{1}+\alpha _{2}}%
x_{0}^{2}\right) x_{0}.  \label{accel}
\end{equation}

The predictions of the VA are compared to numerical findings in Fig. \ref%
{fig3}, where panel (a) shows the numerically obtained period of
oscillations of the separation between centers of the interacting solitons,
in the case of $\beta >0$, as a function of norm $N_{2}$. The numerical data
are obtained using initial condition $\Phi _{0}(x)=\mathrm{sech~}x$, $\Psi =%
\sqrt{N_{2}/2}\mathrm{sech~}\left( \sqrt{N_{2}/2}(x-x_{0})\right) $, with $%
x_{0}=0.1$, other parameters being
\begin{equation}
M_{1}=1,M_{2}^{-1}=0.8,G_{1}=G_{2}=0.9,\Gamma =0.1.  \label{1.25}
\end{equation}%
The comparison of the analytically predicted period of the oscillations of
separation $\Delta \xi $ between the solitons, see Eq. (\ref{osc}), and
ratio $R$ of the amplitude of the shuttle oscillations of the pair as a
whole to the amplitude of the intrinsic oscillations of $\Delta \xi (t)$,
see Eq. (\ref{R}), with their numerically found counterparts attests to good
accuracy of the analytical approximation. In particular, large values of $R$
explain why the two solitons seem overlapping in Figs. \ref{fig1}(b) and \ref%
{fig2}.
\begin{figure}[h]
\begin{center}
\includegraphics[height=3.5cm]{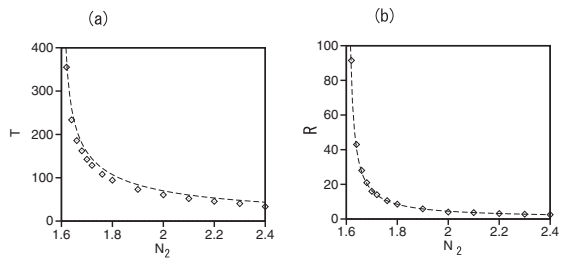}
\end{center}
\caption{(a) Rhombuses represent the numerically obtained period of
oscillations of separation $\Delta \protect\xi $ between the interacting
solitons, as a function of norm $N_{2}$ for $N_{1}=2$, and other parameters
taken as per Eq. (\protect\ref{1.25}). The dashed line shows the analytical
prediction provided by Eq. (\protect\ref{osc}), i.e., $T=2\protect\pi /%
\protect\sqrt{\protect\alpha \protect\beta }$. (b) Rhombuses: the
numerically evaluated ratio $R$ of the oscillation amplitudes of the mean
position of the soliton pair, $\Xi $, and separation $\Delta \protect\xi $
between them. The dashed line shows the respective analytical approximation
given by Eq. (\protect\ref{R}).}
\label{fig3}
\end{figure}

Systematic results for the co-accelerating motion of the pair of solitons at
$\beta =0$ are displayed in Fig. \ref{fig4}. Panel (a) shows a typical
example of numerically generated trajectories of their centers, for
parameters
\begin{equation}
M_{1}=M_{2}^{-1}=1,G_{1}=0.9,G_{2}=1.1,\Gamma =0.1,N_{1}=N_{2}=2,
\label{NN=2}
\end{equation}%
and initial separation $x_{0}=0.5$. Further, the dependence of the
numerically identified acceleration on initial separation $x_{0}$, and its
comparison with the analytical prediction produced by Eq. (\ref{accel}) are
displayed in Fig. \ref{fig4}(b). The presence of the maximum in the
dependence $\mathrm{a}\left( x_{0}\right) $ (at point $x_{0}=1.05$) is
explained by the fact that the interaction force vanishes both at $x_{0}=0$
and at $x_{0}\rightarrow \infty $.

\subsection{The co-accelerating motion of the envelope soliton pair in the
presence of gravity}

Because the gravity also imparts acceleration to matter-wave solitons \cite%
{Modugno}, a natural extension of the above analysis is to add the gravity
potential, $-fx$, with strength $f$, to the system of Eqs. (\ref{PHI}), (\ref%
{PSI}):%
\begin{eqnarray}
i\frac{\partial \Phi }{\partial t} &=&-\frac{1}{2M_{1}}\frac{\partial
^{2}\Phi }{\partial x^{2}}-(G_{1}|\Phi |^{2}+\Gamma |\Psi |^{2}+fx)\Phi ,
\label{grav1} \\
i\frac{\partial \Psi }{\partial t} &=&\frac{1}{2M_{2}}\frac{\partial
^{2}\Psi }{\partial x^{2}}-(-G_{2}|\Psi |^{2}+\Gamma |\Phi |^{2}+fx)\Psi .
\label{grav2}
\end{eqnarray}%
We stress that, while the derivation of Eqs. (\ref{PHI}), (\ref{PSI}) and (%
\ref{grav1}), (\ref{grav2}) from the underlying GPE system (\ref{cos}),
including the OL potential (and the gravity potential, in the present
context), may generate the negative effective dynamical mass, $-M_{2}$,
gravity masses of the solitons represented by envelope wave functions $\Phi $
and $\Psi $ remains normal (positive), therefore the gravity potentials have
the same sign in Eqs. (\ref{grav1}) and (\ref{grav2}).

The VA outlined above can be readily extended to include the gravity, which
yields the following modification of Eq. (\ref{xi1}):
\begin{gather}
\frac{d^{2}\xi _{1}}{dt^{2}}=\frac{N_{2}}{M_{1}}\alpha \exp \left[ -\frac{%
2\alpha _{1}\alpha _{2}}{\alpha _{1}+\alpha _{2}}(\xi _{1}-\xi _{2})^{2}%
\right] (\xi _{2}-\xi _{1})+\frac{f}{M_{1}},  \label{fxi1} \\
\frac{d^{2}\xi _{2}}{dt^{2}}=\frac{N_{1}}{M_{2}}\alpha \exp \left[ -\frac{%
2\alpha _{1}\alpha _{2}}{\alpha _{1}+\alpha _{2}}(\xi _{1}-\xi _{2})^{2}%
\right] (\xi _{2}-\xi _{1})-\frac{f}{M_{2}},  \label{fxi2}
\end{gather}%
and respective changes in Eqs. (\ref{Delta_xi}) and (\ref{Xi}):%
\begin{gather}
\frac{d^{2}\Delta \xi }{dt^{2}}=-\alpha \beta \exp \left[ -\frac{2\alpha
_{1}\alpha _{2}}{\alpha _{1}+\alpha _{2}}(\Delta \xi )^{2}\right] \Delta \xi
-\left( \frac{1}{M_{2}}+\frac{1}{M_{1}}\right) f,  \label{fDelta_xi} \\
\frac{d^{2}\Xi }{dt^{2}}=\frac{\alpha }{2}\left( \frac{N_{2}}{M_{1}}+\frac{%
N_{1}}{M_{2}}\right) \exp \left[ -\frac{2\alpha _{1}\alpha _{2}}{\alpha
_{1}+\alpha _{2}}(\Delta \xi )^{2}\right] \Delta \xi   \notag \\
+\frac{1}{2}\left( \frac{1}{M_{1}}-\frac{1}{M_{2}}\right) f.  \label{fXi}
\end{gather}%
\begin{figure}[h]
\begin{center}
\includegraphics[height=3.5cm]{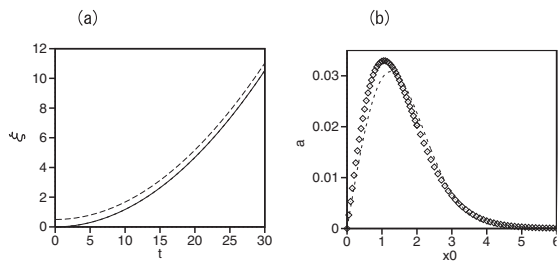}
\end{center}
\caption{(a) The continuous and dashed curves display a typical example of
numerically generated trajectories of centers of the interacting solitons,
in the case of their co-accelerating motion. The respective parameters are
given in Eq. (\protect\ref{NN=2}), with $x_{0}=0.5$. Both trajectiries are
close to parabolas, with acceleration $\mathrm{a}=0.0234$. (b) Rhombuses
represent numerically found values of the co-acceleration, $\mathrm{a}$, as
a function of $x_{0}$, for the same parameters, the dashed curve showing the
analytical approximation given by Eq. (\protect\ref{accel}).}
\label{fig4}
\end{figure}
\

\begin{figure}[h]
\begin{center}
\includegraphics[height=3.5cm]{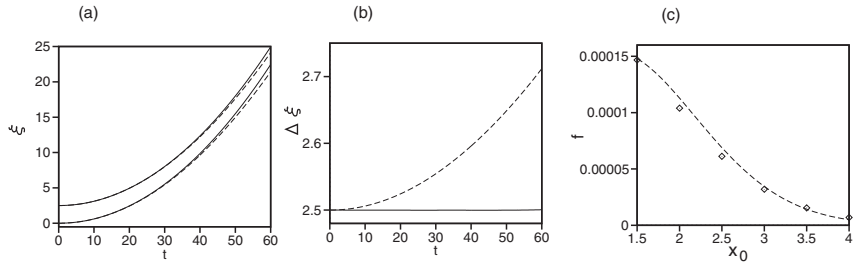}
\end{center}
\caption{(a) Numerically obtained trajectories of centers $\protect\xi _{1}$
and $\protect\xi _{2}$ of the interacting solitons in the absence and
presence of the gravity, \textit{viz}., at $f=0$ and $f=6.1\cdot 10^{-5}$
(the dashed and continuous lines, respectively) for parameter given by Eq. (%
\protect\ref{with_f}). (b) The evolution of the separation between the
solitons, $\Delta \protect\xi $, in the same cases. (c) Rhombuses represent
numerically found values of the gravity strength, $f$, corresponding to
stable pairs of co-accelerating solitons, as a function of the initial
separation between them, $x_{0}$. The dashed curve is the respective
analytical prediction, produced by Eq. (\protect\ref{f}).}
\label{fig5}
\end{figure}

The gravity may be used to compensate the splitting force in the case of $%
\beta <0$, and thus switch the splitting regime into the co-acceleration. A
straightforward analysis demonstrates that the balance between the
interaction and gravity forces produces a stable co-acceleration regime at $%
\Delta \xi >1.05$, where the slope of the curve in Fig. \ref{fig4}(b) is
negative. As a typical example, Figs. \ref{fig5}(a) and (b) show
trajectories of centers of the co-accelerating solitons, and the evolution
of the separation between them at $f=0$ and $f=6.1\cdot 10^{-5}$, for
parameters
\begin{equation}
M_{1}=1,M_{2}^{-1}=1.01,G_{1}=0.9,G_{2}=1.11,\Gamma =0.1,N_{1}=N_{2}=2,
\label{with_f}
\end{equation}%
and initial separation $x_{0}=2.5$. It is seen that the gravity maintains
the stable co-acceleration. On the other hand, in the absence of the
gravity, the solitons exhibit, in Fig. \ref{fig5}(b), slow separation.

Further, Fig. \ref{fig5}(c) shows the gravity strength in the stable
co-accelerating pair as a function of the initial separation, $x_{0}$, as
found from numerical data, and compared to the analytical prediction, which
is produced by Eq. (\ref{fDelta_xi}):
\begin{equation}
f=-\frac{M_{1}M_{2}\alpha }{M_{1}+M_{2}}\exp \left( -\frac{2\alpha
_{1}\alpha _{2}}{\alpha _{1}+\alpha _{2}}x_{0}^{2}\right) x_{0}\beta .
\label{f}
\end{equation}%
Note that the inverse relation, $\beta =-\left( M_{1}M_{2}\alpha
x_{0}\right) ^{-1}\left( M_{1}+M_{2}\right) \exp \left[ 2\alpha _{1}\alpha
_{2}\left( \alpha _{1}+\alpha _{2}\right) ^{-1}x_{0}^{2}\right] f$, defines
the value of $\beta $ at which the robust regime of the co-acceleration
occurs, replacing condition $\beta =0$ [see Eq. (\ref{beta})], derived above
in the absence of gravity. Thus, the gravity may be used to adjust the
occurrence of the co-acceleration regime, for given values of other
parameters (in particular, $\beta $). The necessary value of $f$ can be
readily tuned by varying the angle, $\theta $, between the vertical axis and
direction of the quasi-one-dimensional waveguide into which the BEC is
loaded: $f=f_{\max }\cos \theta $, where $f_{\max }$ corresponds to the
waveguide oriented parallel to the gravity force.

\subsection{Collisions between pairs of envelope solitons}

\begin{figure}[h]
\begin{center}
\includegraphics[height=3.5cm]{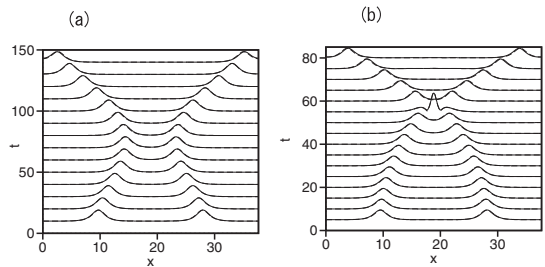}
\end{center}
\caption{Collisions of solitons pairs, generated by initial conditions (%
\protect\ref{x0L}). (a) Pairs performing shuttle motion at parameters given
by Eq. (\protect\ref{shuttle}). (b) Co-accelerating pairs, corresponding to
parameters (\protect\ref{coaccel}). In both panels, profiles of $\left\vert
\Phi (x,t)\right\vert $ and $\left\vert \Psi \left( x,t\right) \right\vert $
strongly overlap.}
\label{fig6}
\end{figure}

Another relevant issue is to consider collisions between the soliton
complexes. Typical examples, produced by simulations of Eqs. (\ref{PHI}) and
(\ref{PSI}) (in the absence of the gravity), are displayed in Fig. \ref{fig6}%
, starting from the corresponding input,
\begin{gather}
\Phi _{0}(x)=\mathrm{sech}(x-L/4)+\mathrm{sech}(x-3L/4),  \notag \\
\Psi _{0}(x)=\mathrm{sech}(x-L/4-x_{0})+\mathrm{sech}(x-3L/4+x_{0}),  \notag
\\
x_{0}=0.1,~L=12\pi  \label{x0L}
\end{gather}%
(this input implies that the integration-domain's center is located at point
$x=L/2$).

First, for parameters%
\begin{equation}
M_{1}=1,M_{2}^{-1}=0.98,G_{1}=0.9,G_{2}=1.06,\Gamma =0.1,  \label{shuttle}
\end{equation}%
at which the pair of interacting solitons perform the shuttle motion, Fig. %
\ref{fig6}(a) demonstrates that the two pairs collide and bounce back.
Detailed analysis of the numerical data demonstrates that the collision
result in an increase of the separation $\Delta \xi $ in each pair and,
respectively, increase of the amplitude of the shuttle oscillations.

The collision between two soliton pairs which move with the co-acceleration
in opposite directions is displayed in Fig. \ref{fig6}(b), for parameters%
\begin{equation}
M_{1}=1,M_{2}^{-1}=1,G_{1}=0.9,\Gamma =0.1,G_{2}=1.1.  \label{coaccel}
\end{equation}%
In this case the colliding pairs pass through each other and, similar to the
case displayed in Fig. \ref{fig6}(b), the collision results in an increase
of the separation between the interacting solitons in each pair, from $%
\Delta \xi =0.10$ to $\Delta \xi \approx 0.17$. This, in turn, leads to the
increase of the co-acceleration, as per Eq. (\ref{accel}) and Fig. \ref{fig4}%
(b). The enhanced self-acceleration is clearly observed in Fig. \ref{fig6}%
(b).

\section{Co-accelerating soliton pairs in the underlying system with the
optical-lattice potential}

The above considerations, both numerical and analytical, were performed for
Eqs. (\ref{PHI}) and (\ref{PSI}), which govern the evolution of envelope
wave functions $\Phi \left( x,t\right) $ and $\Psi \left( x,t\right) $. It
is also relevant to verify the possibility of the shuttle and
co-accelerating motion of the soliton pairs in the framework of the
underlying GPEs (\ref{cos}), which explicitly include the OL potential and
original physical coefficients, rather than the effective ones produced by
averaging, as given by Eqs. (\ref{meff1}), (\ref{meff2}), and (\ref{geff}).
We also consider the version of Eq. (\ref{cos}) which includes the gravity
potential, similar to Eqs. (\ref{grav1}) and (\ref{grav2}):
\begin{eqnarray}
i\frac{\partial \phi }{\partial t} &=&-\frac{1}{2}\frac{\partial ^{2}\phi }{%
\partial x^{2}}-\left[ g_{1}|\phi |^{2}+\gamma |\psi |^{2}+U_{1}\cos \left(
2\pi x\right) +fx\right] \phi ,  \notag \\
i\frac{\partial \psi }{\partial t} &=&-\frac{1}{2}\frac{\partial ^{2}\psi }{%
\partial x^{2}}-\left[ \gamma |\phi |^{2}-g_{2}|\psi |^{2}+U_{2}\cos \left(
2\pi x\right) +fx\right] \psi .  \label{fcos}
\end{eqnarray}%
Note that gravity strength $f$ in Eq. (\ref{fcos}) is the same as in Eqs. (%
\ref{PHI}) and (\ref{PSI}), because the derivation of the latter equations
from the former ones does not involve rescaling of variables $t$ and $x$.

Equations (\ref{cos}) and (\ref{fcos}) were solved numerically with various
initial conditions. In particular, input
\begin{gather}
\phi _{0}(x)=A~\frac{1+2a\cos \left( \pi x\right) }{\sqrt{1+2a^{2}}}\mathrm{%
sech}(A(x-0.5)),\;  \notag \\
\psi _{0}(x)=\sqrt{2}B\cos \left( \pi x\right) \mathrm{sech}(Bx),
\label{phipsi}
\end{gather}%
with $a$ defined as per Eq. (\ref{a}), is suggested by the above
approximations (\ref{phi}) and (\ref{psi}) for the wave functions.

\begin{figure}[h]
\begin{center}
\includegraphics[height=3.8cm]{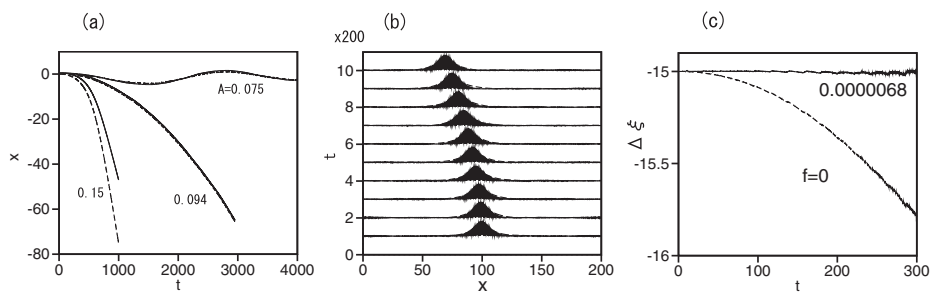}
\end{center}
\caption{(a) The continuous and dashed lines represent trajectories of the
motion of centers of components $\protect\phi $ and $\protect\psi $,
produced by simulations of Eq. (\protect\ref{cos}) with $U_{1}=0$, $U_{2}=8$
and input (\protect\ref{phipsi}), in which $B=0.15$ is fixed, while values $%
A=0.075$, $0.094$, and $0.150$ are adopted for the other amplitude. (b) The
evolution of (virtually coinciding) $|\protect\phi \left( x,t\right) |$ and $%
|\protect\psi \left( x,t\right) |$ for $A=0.094$. (c) The evolution of
separation $\Delta \protect\xi $ between the interacting components at $%
A=0.110$ and $x_{0}=15$, in the absence of the gravity (the dashed line),
and in presence of the gravity potential with strength $f=-6.8\cdot 10^{-6}$
(the solid line), as produced by simulations of Eqs. (\protect\ref{fcos}),
which include both the OL and gravity potentials. Other parameters are fixed
according to Eq. (\protect\ref{1.11}).}
\label{fig7}
\end{figure}

First, Fig. \ref{fig7} displays numerical results obtained by simulations of
Eq. (\ref{cos}) with initial conditions (\ref{phipsi}) in the case of $%
U_{1}=0$ and $U_{2}=8$, that is, assuming that the OL potential acts only on
the $\psi $ component (typical results for the setting with $U_{1}=U_{2}$
are displayed below). Figure \ref{fig7}(a) shows trajectories of the motion
of centers of the two components for parameters

\begin{equation}
g_{1}=0.8,\gamma =0.2,g_{2}=\frac{2}{3}\left( \gamma +\frac{2\pi ^{2}-U_{2}}{%
U_{2}}\right) \equiv 1.11  \label{1.11}
\end{equation}%
in Eq. (\ref{cos}), and amplitude $B=0.15$ in Eq. (\ref{phipsi}), while
amplitude $A$ is varied, taking values $A=0.075$, $0.094$, and $0.15$ [the
particular choice of $g_{2}$ in Eq. (\ref{1.11}) is made to facilitate the
prediction of the value of $A$ at which the co-accelerating regime may be
expected, see Eq. (\ref{A0}) below]. The choice of the smallest amplitude, $%
A=0.075$, gives rise to the shuttle motion, while the largest amplitude, $%
A=0.15$, leads to splitting of the soliton pair. The regime of the robust
co-acceleration of the two solitons, which keep a constant separation
between themselves, is found at
\begin{equation}
A=A_{0}^{\mathrm{(num)}}\approx 0.094.  \label{num}
\end{equation}

The analytical approximation, based on the above condition $\beta =0$ [see
Eq. (\ref{beta})], with effective mass and interaction coefficients
calculated as per Eqs. (\ref{meff1}), (\ref{meff2}), and (\ref{geff}),
yields the value
\begin{equation}
A_{0}=BU_{2}/(2\pi ^{2}-U_{2})=0.102,  \label{A0}
\end{equation}%
at which the co-accelerating regime is predicted, the respective negative
effective mass being $-M_{2}=-A_{0}/B=-0.6266$. A difference ($\Delta
A_{0}/A_{0}\approx 0.08$) of the predicted value (\ref{A0}) from its
numerical counterpart (\ref{num}) is explained by deviation of the
analytical approximations (\ref{meff1}) and (\ref{meff2}) from numerically
exact values, and also by effects of the emission of radiation from the
solitons moving through the periodic potential.

Figure \ref{fig7}(b) shows the evolution of wave functions in the
co-accelerating pair, in terms of $|\phi \left( x,t\right) |$ and $|\psi
\left( x,t\right) |$, at point (\ref{num}). The pair of solitons are
traveling to left, under the action of the attraction between them, because
the positive-mass $\phi $ soliton is initially set to the right of the
negative-mass one in the $\psi $ component. If the initial configuration is
reversed, the pair moves to right. At $A>0.094$, the pair splits because the
negative-mass soliton runs to left with a larger acceleration than the
positive-mass one is able to develop. On the other hand, at $A<0.094$ the
positive-mass soliton overtakes the negative-mass one and passes it, which
leads to reversal of the direction of motion, inducing the shuttle regime.

The effect of the gravity potential, added to Eq. (\ref{fcos}), is displayed
in Fig.~\ref{fig7}(c). It shows the evolution of separation $\Delta \xi $
between solitons' centers for $A=0.11$ in input (\ref{phipsi}), with initial
separation $x_{0}=15$. According to the above findings, in the absence of
the gravity the pair should split in this case, because amplitude $A$
exceeds the respective critical value, $A_{0}=0.094$. This is indeed
demonstrated by the dashed curve in Fig.~\ref{fig7}(c). On the other hand,
the solid curve shows that the application of gravity with $f=-6.8\cdot
10^{-6}$ offsets the splitting force and creates a co-accelerating pair with
a virtually constant separation, cf. Fig. \ref{fig5}(b).

\begin{figure}[h]
\begin{center}
\includegraphics[height=3.8cm]{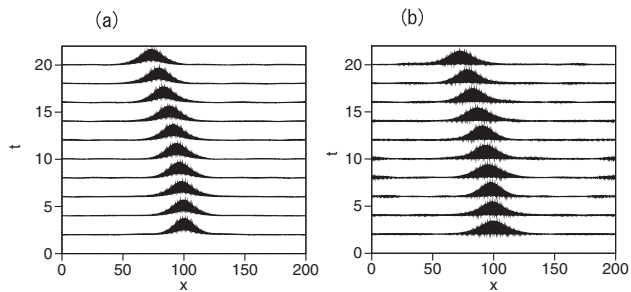}
\end{center}
\caption{The evolution of $|\protect\phi \left( x,t\right) |$ (a) and $|%
\protect\psi \left( x,t\right) |$ (b), produced by simulations of Eq. (%
\protect\ref{cos}) with $g_{1}=0.348$, $g_{2}=0.368$, $\protect\gamma =0.05$%
, and $U_{1}=U_{2}=13$, and initial conditions (\protect\ref{phipsi}) with $%
A=0.103$ and $B=0.15$. The robust regime of the co-acceleration of the
interacting positive ($\protect\phi $)- and negative ($\protect\psi $)-mass
solitons is observed.}
\label{fig8}
\end{figure}

A typical example of the robust co-acceleration regime found in the system
with equal amplitudes of the OL potential acting on both components, \textit{%
viz}., $U_{1}=U_{2}=13$, is displayed in Fig. \ref{fig8}. In this case, the
amplitudes of input (\ref{phipsi}) are $A=0.103$ and $B=0.15$, and the
self-interaction coefficients are taken as $%
g_{1}=(1+2a^{2})^{2}/(1+12a^{2}+6a^{4})\left[ 1/M_{1}-\gamma
(1+2a^{2})/(1+2a^{2}+2a)\right] \equiv 0.348$ and $g_{2}=(2/3)\left[
1/M_{2}+\gamma (1+2a^{2})/(1+2a^{2}+2a)\right] \equiv 0.368$, for $\gamma
=0.05$, where $a$ is defined by Eq. (\ref{a}). In the framework of the above
analytical approximation, these parameters predict the co-accelerating
motion at $\beta =0$ [see Eq. (\ref{beta})], which amounts to the value of
the amplitude $A_{0}=BM_{2}/M_{1}=0.158$. It is essentially larger than the
numerically found value, $A_{0}^{\mathrm{(num)}}\approx 0.103$, at which the
co-acceleration is observed in Fig. \ref{fig8}, i.e., in this case, with the
strong OL potential, the simple analytical approximation produces only
qualitatively correct predictions.

Lastly, while the dynamical regimes of the shuttle motion and
co-acceleration produced by Eqs. (\ref{PHI}) and (\ref{PSI}), or (\ref{grav1}%
) and (\ref{grav2}), may persist indefinitely long, the motion of the
solitons across the OL in the framework of Eqs. (\ref{cos}) and (\ref{fcos})
is accompanied by weak radiation losses, which may be seen as tiny
perturbations in Figs. \ref{fig7} and \ref{fig8}. Eventually, these losses
may essentially damage the solitons, but this will happen on a time scale
essentially exceeding an experimentally relevant one.

\section{Conclusion}

The objective of this work is to establish the framework which admits
co-accelerating motion of interacting objects with opposite signs of the
effective mass, using pairs of matter-wave solitons which move against the
background of the OL (optical-lattice) potential. The effective negative
mass of one component is provided by the known property of gap solitons.
Reducing the full system of the GPEs (Gross-Pitaevskii equations), which
includes the OL potential, to equations for slowly varying envelopes,
systematic simulations and the VA (variational approximation) reveal two
generic dynamical regimes, \textit{viz}., spontaneous shuttle oscillations
of the mean position of the soliton pair, in the course of which the
solitons periodically pass through each other, and splitting of the pair.
The robust co-acceleration of the soliton pairs, with a permanent separation
between the constituents, is found at the boundary between these two
regimes. The location of the boundary can be adjusted by dint of the gravity
potential added to the system. The VA accurately predicts all these effects.
Finally, the same dynamical regimes, including the robust co-acceleration,
are directly demonstrated by simulations of the underlying system, which
includes the OL potential and the gravity potential (if any) as well. The
predicted effects can be realized experimentally in two-component atomic
BEC, loaded in a quasi-one-dimensional waveguide combined with the OL, and
the occurrence of the co-acceleration regime can be adjusted by choosing the
angle between the waveguide and gravity direction.

It may be interesting to consider a modification of the model which includes
linear interconversion (Rabi coupling) between the components, which may
help to additionally bind them, cf. Ref. \cite{Merhasin}. A challenging
possibility is to develop a two-dimensional version of the present system
and, accordingly, to study pairs of two-dimensional solitons in the regimes
of co-acceleration and spontaneous shuttle motion.

Finally, it is also relevant to mention that, in addition to the ultracold
atomic gases, exciton-polariton BECs have been experimentally realized in
semiconductor microcavities \cite{microcav,microcav-review}, and predicted
in graphene and similar two-dimensional materials \cite{graphene,Berman}, at
temperatures exceeding those necessary for the condensation of bosonic gases
by eight or nine orders of magnitude. Polariton solitons have also been
created in microcavities \cite{microcav-soliton}, and it is expected that
they may exist in graphene-like settings as well \cite%
{graphene-predic-soliton}. These findings suggest a possibility to create
coupled positive- and negative-mass soliton pairs in polariton BEC. However,
the necessary analysis will be completely different from that reported in
the present paper, as media supporting polaritons are essentially
dissipative, hence a pump must be included too. The latter term (unlike the
simple dissipation) destroys the Galilean invariance, thus making the
consideration of accelerating and shuttle dynamical regimes a challenging
problem, which should be considered elsewhere.

\section*{Acknowledgments}

We appreciate valuable discussions with U. Peschel and Yu. V. Bludov.

\end{document}